# Interdisciplinary Research Collaborations: Evaluation of a Funding Program*

**Nadine Rons**
Vrije Universiteit Brussel (VUB), Pleinlaan 2, B-1050 Brussels, Belgium
Nadine.Rons@vub.ac.be

Nadine Rons (°1969) holds a Master degree in Physics from the Vrije Universiteit Brussel (1991), where she did research in astronomy for four years until 1995. In 1996 she joined the university's Research Coordination unit. Her research interests cover a wide range of topics related to research management, including reliability of evaluation results, and indicators for research performance for different research fields and aggregation levels. Among major developments for the university are a method for the evaluation of research teams per discipline by international peers and an internal allocation model for research funding.

**Innovative ideas are often situated where disciplines meet, and socio-economic problems generally require contributions from several disciplines. Ways to stimulate interdisciplinary research collaborations are therefore an increasing point of attention for science policy. There is concern that 'regular' funding programs, involving advice from disciplinary experts and discipline-bound viewpoints, may not adequately stimulate, select or evaluate this kind of research. This has led to specific policies aimed at interdisciplinary research in many countries. There is however at this moment no generally accepted method to adequately select and evaluate interdisciplinary research. In the vast context of different forms of interdisciplinarity, this paper aims to contribute to the debate on best practices to stimulate and support interdisciplinary research collaborations. It describes the selection procedures and results of a university program supporting networks formed 'bottom up', integrating expertise from different disciplines. The program's recent evaluation indicates that it is successful in selecting and supporting the interdisciplinary synergies aimed for, responding to a need experienced in the field. The analysis further confirms that potential for interdisciplinary collaboration is present in all disciplines.**

## 1. Interdisciplinarity: variety and focus

The term 'interdisciplinarity' is used in relation to many different manifestations of the phenomenon, involving different actors, sectors and interactions. The various kinds of interdisciplinarity have been listed, codified and studied from practical experiences (Klein [1]; Frodeman, Klein and Mitcham [2]; Weingart and Stehr [3]) as well as from a cognitive point of view (Derry, Schunn and Gernsbacher [4]). A concise and up to date taxonomy of interdisciplinarity in general is given by Klein [5]. Looking more specifically at interdisciplinarity in research, the need for crossing the boundaries of research disciplines is a natural phenomenon. The reality that one wishes to study and describe remains a complex, 'interdisciplinary' combination of aspects and properties, regardless of the disciplinary delineations installed by man. From this point of view, combinations of knowledge originating from several such 'artificially' separated disciplines are equally valuable in their potential to advance science as those stemming from a same discipline. Efforts to categorize and order human knowledge have accompanied its development since the beginning. The present division of scientific knowledge in disciplines (physics, chemistry, ...) emerged about two centuries ago as

* Based on the paper presented at the Sixth International Conference on Webometrics, Informetrics and Scientometrics & Eleventh COLLNET Meeting held in Mysore, India, 19-22 October 2010

researchers got organized in more focused communities, being confronted with growing amounts of scientific information produced and communicated. These large disciplines remained relatively stable and became institutionalized in higher education structures, where they have been shaping knowledge production, funding and distribution. Not long after the introduction of disciplinary boundaries, these were subject to opposition from the Unity of Science Movement, striving for one single scientific language (Neurath, Carnap and Morris [6]), and a seminal study on interdisciplinarity appeared, linked to the first international conference on the subject sponsored by the OECD (Apostel et al. [7]). Presently, the idea that the current structural disciplinary organization hampers potentially valuable and innovative interdisciplinary interactions is widely accepted. Studies of science defend interdisciplinarity in research for its contributions to scholarship and society (Nissani [8]) and point to knowledge of interdisciplinary 'structural holes' as a potential competitive advantage to researchers in view of creativity and discovery (Chen et al. [9]), and adaptations in organization and management are proposed to facilitate interdisciplinary research (COSEPUP [10]). Also political and socio-economic priority areas have stimulated interdisciplinary research by requiring contributions from different disciplines (e.g. nanotechnology, climate research), and by making the additional 'transdisciplinary' step towards application and innovation (Barry, Born and Weszkalnys [11]). The fact that the amount of knowledge continues to grow (Persson [12]) is likely to further enhance the creation of more, and more specialized research communities and thereby also the need to encourage and facilitate communication and collaboration between them. In the present era of information technology, the growing scientific literature is captured and linked in global as well as thematic international publication and citation databases. Already in the pioneering paper leading to the creation of the current Web of Knowledge's predecessor, the need for interdisciplinary communication was present in the possible utility attached to the proposed index. The new index would "bring together material that would never be collated by the usual subject indexing", in "an association-of-ideas index" across the subject headings of the then conventional indices, serving "the subject approach of the scientist who seeks information" (Garfield [13]). This idea is present more explicitly in the paper's successor that emphasizes the importance of the new citation index "in view of the accelerating tempo of modern interdisciplinary scientific research" (Garfield [14]). Especially in the broadest publication and citation databases created since, i.e. the Web of Knowledge and Scopus, the disciplinary labels provide a basis for bibliometric instruments to investigate disciplinary structures, evolutions in the organization of science, and different kinds of interdisciplinarity in research. Porter and Chubin [15] were among the first to investigate cross-disciplinary research in this way, proposing citations by and to other disciplines as an indicator. Rinia et al. [16] demonstrated the larger time lag of such cross-disciplinary citations and used them to study the rate of knowledge transfer between fields at macro-level, visible in their 'natural' shares of cross-disciplinary activity (Rinia et al. [17]), while Porter and Rafols [18] studied their evolution over time. An example at micro-level is the study by Porter et al. [19] on interdisciplinarity of samples of researchers, using discipline-based measures for integration and specialization, later used to evaluate the National Academies Keck Futures Initiative (NAKFI) facilitating interdisciplinary research in the United States (Porter, Roessner and Heberger [20]). Bordons, Morillo and Gómez [21] provide a review of main objectives and methodologies in the analysis of cross-disciplinary research from the specific point of view of quantitative studies of science. Ultimately, among the extensive literature in informetrics, research on interdisciplinarity in research as a self-standing field of science and research policy has in itself become an object of bibliometric study (Braun and Schubert [22]). Conceptual categorizations of interdisciplinary research, created from a theoretical approach (Klein [5]), are rarely used in empirical analyses. Huutoniemi et al. [23] propose an epistemologically grounded conceptual framework for identification and categorization in practice of the very different kinds of interdisciplinarity that can be found in research documents, such as research proposals. Funding programs and research policies can indeed address or give rise to various kinds of interdisciplinary research, depending on their criteria and other specificities. Examples of interdisciplinary research policies and practices from eight different countries were studied for the research project 'Challenging Knowledge and Disciplinary Boundaries through Integrative Research Methods in the Social Sciences and Humanities', funded by the European Commission under Framework 6 (CIT2-CT-2004-506013). Many of the policies and instruments aimed at interdisciplinary research have a 'top down', thematic approach, such as the two UK cross-research council funding programs investigated by Griffin et al. [24]. Even though many programs and policies were put in place, there are not yet any generally accepted methodologies to adequately select and evaluate interdisciplinary research. In a special issue of the journal Research Evaluation that was dedicated to the assessment of interdisciplinary research, the heterogeneity of criteria figuring in the different empirical studies clearly indicated the need for further exploration (Laudel and Origgi [25]). Porter et al. [26] list a wide range of factors impacting interdisciplinary research and suggest a series of further research opportunities, including aspects of proposal reviewing. In a literature review on principles and difficulties faced in the evaluation of interdisciplinary and transdisciplinary research, Klein [27] proposes a framework of seven generic evaluation principles.

This paper aims to contribute to the ongoing search for best practices, by presenting the results of the evaluation of a funding program dedicated to the support of 'bottom up' research synergies, integrating expertise of teams from different disciplines.

## 2. Program procedures and criteria

The ‘Horizontal Research Actions’ (HOA) program was set up at the Vrije Universiteit Brussel in 2002, to support research collaborations integrating expertise from different disciplines, around topics proposed by the applicants. Seven calls have been issued since then, until the program was evaluated in 2009 (Table 1). Criteria for ex ante evaluation of the applications concern both the topic (interdisciplinary and innovative character, completeness and added value of the collaboration, importance for science and society) and the strength of the network partners. The program indeed aims to support joint initiatives of excellent teams, expecting that strong disciplinary performance is required for successful interdisciplinary collaborations. Selected projects are funded initially for two years, in the majority of cases extended until four years after mid-term evaluation.

Table 1. Key figures of the 'Horizontal Research Actions' program, 2002-2009

| Key figure | Quantity |
|---|---|
| calls | 7 |
| faculties involved | 8 (all) |
| departments involved | 60 (from a total of 81, or 74%) |
| (co)applicants involved (distinct names) | 166 |
| funded applications | 23 |
| unfunded applications | 38 |
| standard budget per project | 320.000 EUR over 4 years |

Crucial in the ex ante evaluation by peers of interdisciplinary research initiatives is the composition and functioning of the evaluation committee (Langfeldt [28]; Lamont, Mallard and Guetzkow [29]). Several aspects may hamper a correct evaluation of the quality of the proposed research, such as a partial coverage of the whole of the fields concerned, conflicting assumptions regarding quality and discipline related bias. For the HOA program, the university’s tradition in peer review evaluations was extended with a new form of assessment. The evaluation committee is not composed of experts from each of the particular fields of the network partners, but consists of the members of the Board of the Research Council. As a committee, these combine the broad scope, open attitude to different standards and coherence required for a comparative assessment of the interdisciplinary applications. Linked to their function and experience on the Board, all committee members have acquired a good overview of the expertise and performance of the university’s teams in the large domains that they represent (i.e. one of the faculty clusters ‘Social Sciences and Humanities’, ‘Basic, Natural and Applied Sciences’ or ‘Biomedical Sciences’, corresponding to the domains of three permanent committees of the Research Council) and even beyond. Their views are expected to surpass disciplinary perspectives and stand above conservative disciplinary forces.

The Board of the Research Council selects applications in two phases (Table 2). In a first pre-selection phase, the members of the Board use knowledge of performance levels previously demonstrated by the teams, and concentrate on the extent to which the applications meet the program's aims related to the interdisciplinary collaboration, including the extent to which an all-encompassing expertise is offered in the proposed theme and the potential added value of the project for science and society. These criteria correspond to how panel members for the evaluation of multidisciplinary fellowship applications for themselves define a good interdisciplinary proposal, i.e. having the capacity to achieve the stated purpose, combining breath and originality with a mastering of the research tools from the different disciplines (Lamont, Mallard and Guetzkow [29]). A good selection on this basis also ensures that the funded projects hold the necessary features to later score on the three fundamental grounds suggested by researchers from interdisciplinary research institutes to examine the quality of interdisciplinary research outcomes, i.e. (1) consistency with multiple disciplinary antecedents, (2) balance in weaving together perspectives, and (3) effectiveness in advancing understanding (Feller [30]). For each application (at least) two reviews are collected: one by a member of the Board and one by an additional experienced reviewer, both not involved in the applications. The members of the Board present their own reviews as well as those of their 'co-reviewers' to the Board, which makes the pre-selection.

In a second phase, the applicants of the pre-selected projects are invited to present and defend their project before the Board. After each presentation, remaining questions and specific points of attention are discussed. The HOA funding is intended to be spent primarily on one or more researchers embodying the integration of expertise from the different disciplines. The way the teams plan to fill in these positions is an important point of attention in this second phase of the evaluation. Also, where long term potential is present, the support by the Research Council is expected to lead to the attraction of external funding to ensure the continuation of the network, and the teams' strategies to this respect are another point of attention. After all presentations and discussions, the Board formulates its final advice for selection to the Research Council.

Table 2. Phases and procedures of the Horizontal Research Actions program

| *Phases & actors* | *Procedures & criteria* |
|---|---|
| **Application** | |
| Applicants | Submission according to the call's requirements:<br>- Title/theme of the project/network<br>- Leading applicant and co-applicant names and affiliations<br>- Participating teams, expertise brought in by each and elements demonstrating the international recognition thereof<br>- Short project description (max. 3 A4): working hypothesis, consecutive project phases and timing, theme's relevance in context<br>- Short description and motivation of the budget applied for, per year and type of costs (operational, equipment, personnel) + reporting, if present, contributing external funding or collaborations |
| **Pre-selection** | |
| Board of the Research Council (Board) | Designation of reviewers:<br>- Two reviewers per application, among which one member of the Board<br>- Experienced evaluators<br>- Not involved in the applications |
| Reviewers | Assessment according to the program's criteria:<br>- Multidisciplinary and innovative character of the scientific project<br>- Added value at scientific, societal, philosophical, socio-economic, juridical, technical or other level<br>- Added value through interdisciplinary collaboration<br>- All-encompassing expertise offered concerning the theme<br>- Planned participations in international and European networks<br>- Added value of the one or more full time equivalent researcher(s) financed by the project |
| Board | - Presentation of the reviews to the Board by the member(s) of the Board and discussion<br>- Pre-selection of applications<br>- Invitation of the applicants of the pre-selected projects for a brief oral presentation, with instructions regarding timing and content |
| **Selection** | |
| Applicants & Board | - Presentation of the pre-selected projects to the Board by the applicants, each followed by a questions and answers session |
| Board | - Concluding discussion and final advice concerning selection |
| Research Council | - Selection of applications based on the Board's advice |

## 3. Program evaluation: method and phases

The 'Horizontal Research Actions' program started out with a modest budget that was soon enlarged after it appeared to attract many valuable applications. It was evaluated in 2009, when the first four generations of applications could be followed for three years after the start of funding. The evaluation of the program was conducted by the university's Research Coordination Unit and consisted of three consecutive phases, investigating (I) the degree of interdisciplinarity of the networks and faculty participation, (II) the interdisciplinary scientific output and citation impact, and (III) validation by the networks of the data generated in the previous phases in a short survey (Table 3).

Table 3. Phases of the 'Horizontal Research Actions' (HOA) program evaluation

| *Phases & input* | *Results* |
|---|---|
| **I. Analysis of HOA application data** | |
| - (co)applicants and affiliations (department, faculty, faculty cluster)<br>- application generation | - application profiles regarding **degree of interdisciplinarity** and **faculty participation** |
| **II. Bibliometric study** | |
| - (co)applicants and affiliations for HOA applications of the first four generations | - co-publications and co-citations in the SCIE, SSCI, AHCI and Proceedings indices of the Web of Science and their evolution over time, as measures for **output** and **citation impact** in the themes of the HOA applications |
| **III. Survey** | |
| - (co)applicants of the funded HOA projects that could be followed for at least four full years since HOA start<br>- their output and citation impact based on the Web of Science from phase II | - **validity** of the data and indicators generated in phase II (HOA-relatedness, other HOA-related output or impact)<br>- applicants' **opinions** regarding stimulation and support of interdisciplinary research by the HOA program |

In phase I, the degree of interdisciplinarity of all funded and unfunded networks was analyzed based on the affiliations of the applicants to departments and faculties. A distinction was made between "broad", "medium" and "narrow" interdisciplinary collaborations, respectively joining applicants from different faculty clusters (broad), from different faculties within one cluster (medium), and from different departments within one faculty (narrow). A similar distinction between "big" and "small" interdisciplinarity, standing for predominance of links between distant areas versus close disciplines, was used before by Morillo, Bordons and Gómez [31], based on the terminology from Schmoch et al. [32]. Both divisions are consistent with the terminology from integrative studies discerning the concepts of "narrow" and "broad" interdisciplinarity, where narrow interdisciplinarity refers to interaction between disciplines with comparable methods and paradigms (such as history and literature) and broad interdisciplinarity refers to interaction between disciplines with clearly different paradigms and methods (such as disciplines from sciences and humanities).
In phase II, interdisciplinary output and citation impact were analyzed based on the on line Web of Science, for the first four generations of applications (funded and unfunded), which could be followed for at least three years after start of funding. Of the 36 applications concerned, 4 networks completely situated in 'Social Sciences and Humanities' were excluded from this bibliometric analysis, due to insufficient coverage of such networks. Output was measured by "co-publications", defined as joint publications by applicants from different departments. Bordons et al. [33] used a similar output measure for the evaluation of the Multidisciplinary Research Programme at the Universidad Complutense de Madrid. Some previously defined indicators for interdisciplinary output were not used in this specific context aimed at theme-based synergy and integration. Interdisciplinarity of references in published documents (Porter and Chubin [15]; Porter et al. [19]) was not used as it may to a large extent detect output resulting from interdisciplinary applications besides synergies, and may partly reflect the 'natural' interdisciplinary activity in the applicants' domains. The latter is also the reason why the breath of subject categories associated to the journals in which the output is published (Porter, Roessner, and Heberger [20]) was not used. Citation impact was measured by "co-citations", defined as publications citing applicants from different departments. These were obtained by combining the results of 'Cited Reference Search' operations performed for each of the applicants in a network. Out of scope of the evaluation of this recent funding program are impacts and effects that may arise in the long term towards applications and policy.

In phase III, a survey directed to the five earliest funded networks that could be followed for at least four years after start of funding, sought confirmation of the validity and adequacy of the data and indicators generated in phase II, as well as the applicants' opinions on a limited set of questions.

## 4. Program evaluation: results

### 4.1. Degree of interdisciplinarity and faculty participation

All eight of the university's faculties are represented among the applicants. The faculties' shares of (co)applicants are well correlated with the faculties' shares of potential (co)applicants, which are the leading academic staff (Figure 1). This shows that a potential for interdisciplinary collaboration and a need for specific funding are present in all faculties, and that no faculty in particular is much more predestined than another to engage in interdisciplinary research. It nevertheless can be observed that the larger faculties tend to be over-represented among the applicants ('Medicine & Pharmacy' and 'Science & Bio-Engineering Sciences'), and the smaller faculties under-represented (in particular 'Law & Criminology', 'Economics and Social & Political Sciences' and 'Arts & Philosophy'). A possible explanation may be a scale effect, where a faculty's larger size creates a better environment from which to apply and participate in the program. Another explanation may be found in the 'natural' shares of cross-disciplinary activity of the disciplines themselves, as observed in bibliometric measures (Rinia et al. [17]).

At the level of departments, 74% are represented among the applicants in the period 2002-2009, ranging from about half to all of the departments per faculty. Looking more closely at the numbers of applicants per department, these appear to be distributed very unevenly over the departments within each of the faculties (Figure 2). Analogous to the discussion on the distribution over faculties, size and the nature of the research area may play a stronger role on this smaller scale, with 'Applied Biological Sciences' and 'Medical Imaging and Physical Sciences' emerging as the departments with the highest number of applicants.

Figure 1. Faculty share of (co)applicants vs. potential (co)applicants

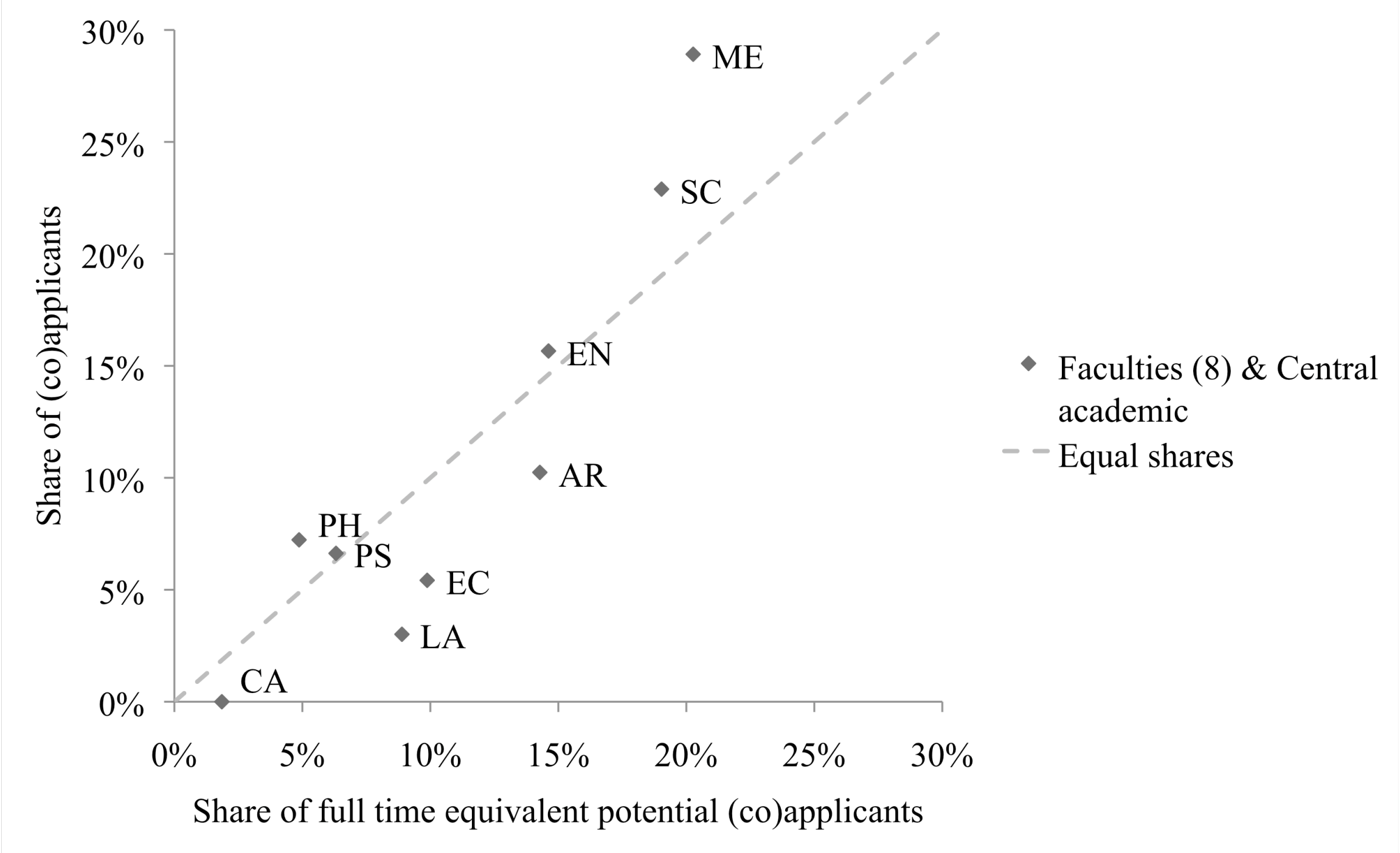

Legend: Faculties: **ME**dicine and pharmacy, **SC**ience and bio-engineering sciences, **EN**gineering sciences, **AR**ts and philosophy, **EC**onomics, social and political sciences, **LA**w and criminology, **PH**ysical education and physiotherapy, **PS**ychology and educational science; **CA**: Central academic; linear correlation coefficient r=0,910 for N=9 (i.e. 8 faculties + central academic staff).

Figure 2. Distribution of (co)applicants over departments, per faculty

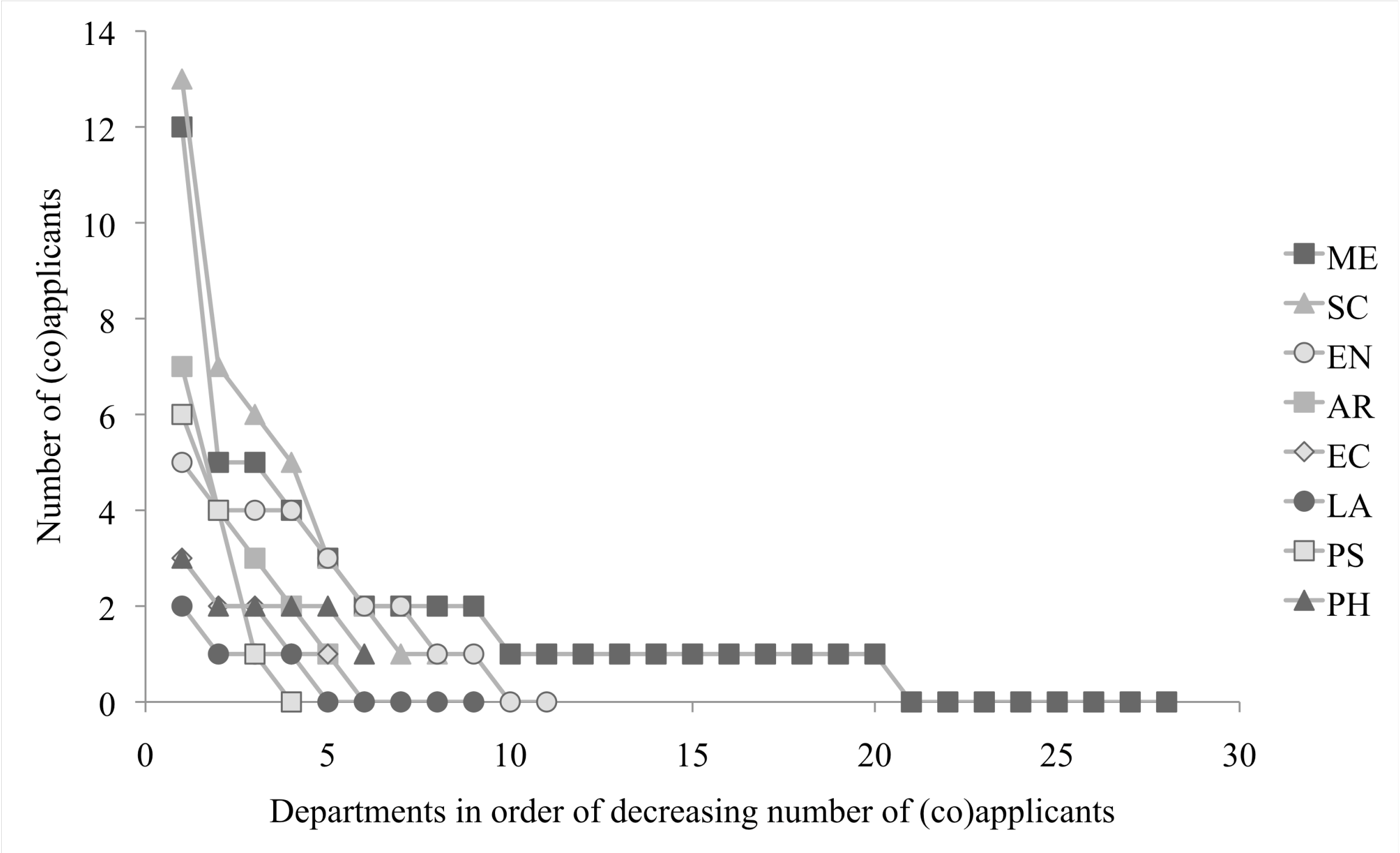

Legend: see Figure 1.

Looking at the composition of the networks, the combinations of departments link each of the eight faculties to at least five others. This indicates a wide potential for the broader types of interdisciplinary collaboration between teams of different faculties. Funded and unfunded proposals show similar network compositions. Both join on average four applicants from three departments, for a majority forming broad interdisciplinary collaborations (Table 4).

Table 4. Network composition of 'Horizontal Research Actions' applications, 2002-2009

| **Network features** | **Funded** | **Unfunded** |
|---|---|---|
| number of applications | 23 | 38 |
| average number of (co)applicants per application | 3,8 | 4,1 |
| average number of departments per application | 2,8 | 3,2 |
| % of networks with broad interdisciplinarity | 65% | 71% |
| % of networks with medium interdisciplinarity | 13% | 16% |
| % of networks with narrow interdisciplinarity | 22% | 13% |

### 4.2. Network activity and impact profiles

Based on their bibliometric interdisciplinary output and citation impact, the following profiles were distinguished among the applications, funded and unfunded:

A. "Newly activated": Newly activated collaborations generating output as well as citation impact. This category was present in particular among the funded applications, of which it contained half, while it was one of the smallest categories among the unfunded applications.

B. "Previously active": Continued previously active collaborations that were already generating output as well as citation impact before the year of application. This category contained about a quarter of the applications, both among the funded and unfunded.

C. "Output only": Newly activated collaborations generating output but not yet citation impact. This was overall the smallest category along with category D.

D. "Citation impact only": Networks not visible in output since application, yet generating joint citation impact. This was overall the smallest category along with category C.

E. "Not visible": Networks not visible in output, nor citation impact. This category was present in particular among the unfunded applications, of which it contained about half, while it was one of the smallest categories among the funded applications.

The shares of the different profiles among funded and unfunded networks (Figure 3) reflect that the program was able to attract promising interdisciplinary applications, which could not all be funded. The program's success in selecting and supporting interdisciplinary research is demonstrated by the fact that a large majority of the funded networks generate output and citation impact (profiles A and B), while about half of the unfunded networks show neither (profile E). At the same time, the findings demonstrate that the program leads to activation of about a quarter of the unfunded networks (profiles A and C), which supports the Research Council's observation that more valuable proposals were attracted than could be funded.

Figure 3. Distribution of bibliometric profiles among funded and unfunded applications, 2002-2006

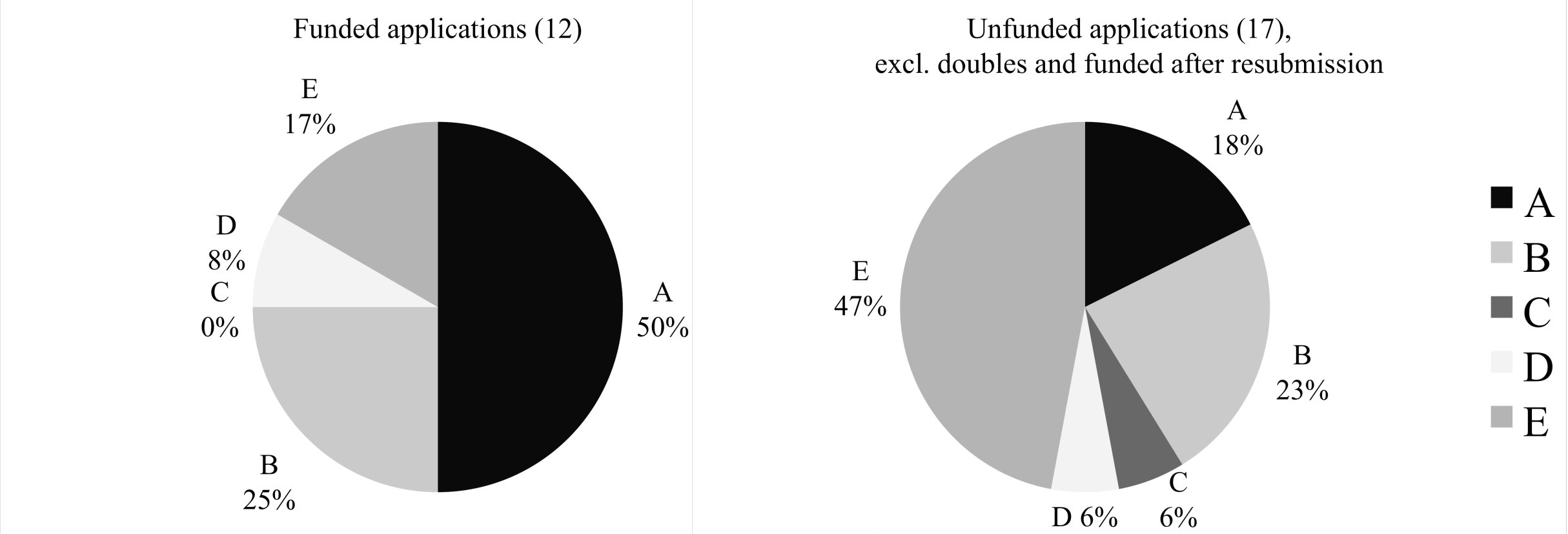

Legend: **A**: "Newly activated", **B**: Previously active", **C**: "Output only", **D**: "Citation impact only", **E**: "Not visible".

Table 5 shows the percentages of bibliometrically visible networks (profiles A to D) for the distinguished degrees of interdisciplinarity. This percentage is highest (83%) among the networks with narrow interdisciplinarity, where more closely related research traditions can be expected to be more easily combined, facilitating the collaboration and consequently its output and impact. The percentage of bibliometrically visible networks among networks with broad interdisciplinarity however is not far behind (62%), considering the potential influence of one network in the limited sample.

Table 5. Percentages of bibliometrically visible networks among funded and unfunded applications, 2002-2006

| **Degree of interdisciplinarity** | **Funded applications (12) + unfunded applications (17, excl. doubles and funded after resubmission)** | **Percentage of bibliometrically visible networks (profiles A to D)** |
|---|---|---|
| Broad | 21 | 62% |
| Medium | 2 | 50% |
| Narrow | 6 | 83% |
| Total | 29 | 66% |

Figure 4 shows the gradual appearance of co-publications for the earliest newly activated networks, which could be followed for at least four years after application. Also co-citations appear gradually (Rons [34]). A comparison of funded and unfunded newly activated networks showed that the added value of funding only becomes clearly visible after about three years, in a more strongly rising output and citation impact.

Figure 4. Co-publications by partners of newly activated interdisciplinary networks

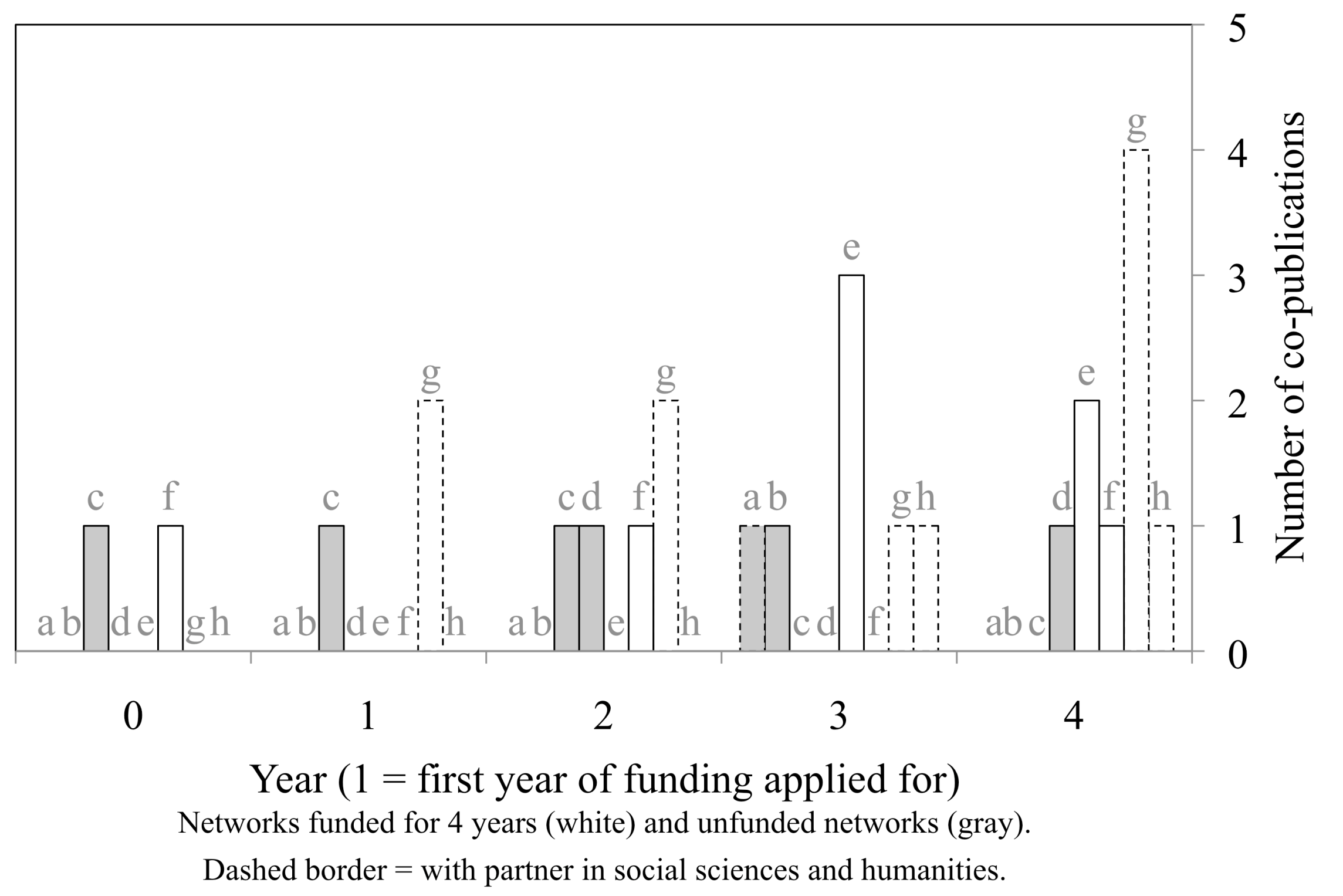

*Publication data sourced from Thomson Reuters Web of Knowledge (formerly referred to as ISI Web of Science)*

**4.3. Data validation and applicants' opinions**

In order to validate the data and indicators generated in phase II, the publications representing output and citation impact were presented to the applicants in a brief survey, inquiring about their synergetic nature, topic-relatedness and completeness. The survey was addressed to the applicants of the five earliest funded networks, all newly activated, for which data could be produced until at least four years after start of funding.

The responses confirmed that the large majority (82%) of traced co-publications concerned a synergy of expertise from different disciplines such as sought by the program, as opposed to the application of expertise from one discipline in the field of another. This shows the program’s success in its aim to support collaborations integrating expertise from different disciplines, which was an important point of attention in the selection procedure.

The responses further confirmed that a majority of the co-publications (89%) and co-citations (64%) generated in phase II is related to the networks’ topics, and that both measures contain the majority of output and impact items of various forms that the applicants relate to their networks. Other categories of output and impact reported were contributions in proceedings, journal publications in Dutch, books in Dutch aimed at a professional audience, PhD's obtained and in preparation, and obtained research funding. These add up to a minority of 30% of output items next to topic-related co-publications in synergy, and to a minority of 11% of impact items next to topic-related co-citations.

The observed high synergetic nature, topic-relatedness and share of output and impact, indicate that the co-publication and co-citation measures used are good representatives of the output and impact of interdisciplinary synergies, and important parameters for their analysis. The absolute numbers of co-publications and co-citations should however be used with caution, as they are not normalized for the very specific interdisciplinary research areas.

In response to the open questions, several respondents confirmed that the HOA program offered a stimulating and necessary funding channel for interdisciplinary research collaborations. Also pointed out several times was the difficulty to combine theoretical and applied research from different domains, doubling the differences in research traditions and enhancing the time needed to bridge them. Similar observations of the challenge that interdisciplinary collaboration implies for researchers trained in different disciplinary practices have been made in other studies. A study by Boix Mansilla, Dillon and Middlebrooks [35] for instance revealed broad curiosity, willingness to embrace risk, disciplinary-rooted open-mindedness, and humility, as the four most prominent dispositions that serve interdisciplinary thinkers, while Krohn [36] points to the ability to critically reassess disciplinary concepts and laws. The magnitude of this challenge implies that assessments of interdisciplinary synergies should take place a sufficiently long time after the start of the collaborations, accounting for a longer start up phase. First results for the earliest networks in the HOA program suggest that this should be after at least three years (Figure 4; Rons [34]).

## 5. Conclusions

In the context of the ongoing search for best practices regarding the assessment of interdisciplinary research, the paper describes a recently evaluated university program designed to support 'bottom up' interdisciplinary networks, integrating expertise from different disciplines. As particular features, the program's two-phased selection procedure involves an evaluation committee consisting of the members of the Board of the Research Council, and an oral presentation of the pre-selected projects by the applicants.

An analysis of the application data indicates that potential for interdisciplinary collaboration and a need for specific funding are present in all faculties and disciplines. A bibliometric study of the output and impact generated by the first generations of applications shows that the program was successful in selecting and supporting the interdisciplinary synergies that it aimed for.